# Electrical Generation of Colour Centres in Hexagonal Boron Nitride

Ivan Zhigulin[1,†], Gyuna Park[2,3,†], Karin Yamamura[1,4], Kenji Watanabe[5], Takashi Taniguchi[6], Milos Toth[1,4], Jonghwan Kim[2,3,7,*], Igor Aharonovich[1,4,*]

1 School of Mathematical and Physical Sciences, University of Technology Sydney, Ultimo, New South Wales 2007, Australia
2 Center for Van der Waals Quantum Solids, Institute for Basic Science (IBS), Pohang, 37673, Republic of Korea
3 Department of Materials Science and Engineering, Pohang University of Science and Technology, Pohang, 37673, Republic of Korea
4 ARC Centre of Excellence for Transformative Meta-Optical Systems, University of Technology Sydney, Ultimo, New South Wales 2007, Australia
5 Research Center for Electronic and Optical Materials, National Institute for Materials Science, 1-1 Namiki, Tsukuba 305-0044, Japan
6 Research Center for Materials Nanoarchitectonics, National Institute for Materials Science, 1-1 Namiki, Tsukuba 305-0044, Japan
7 Department of Physics, Pohang University of Science and Technology, Pohang, 37673, Republic of Korea
† These authors contributed equally to this work.
* To whom correspondence should be addressed: J.K. (jonghwankim@postech.ac.kr) and I.A. (Igor.Aharonovich@uts.edu.au)

**Abstract**

Defects in wide band gap crystals have emerged as a promising platform for hosting colour centres that enable quantum photonic applications. Among these, hexagonal boron nitride (hBN), a van der Waals material, stands out for its ability to be integrated into heterostructures, enabling unconventional charge injection mechanisms that bypass the need for p-n junctions. This advancement allows for the electrical excitation of hBN colour centres deep inside the large hBN bandgap, which has seen rapid progress in recent developments. Here, we fabricate hBN electroluminescence (EL) devices that generate narrowband colour centres suitable for electrical excitation. The colour centres are localised to tunnelling current hotspots within the hBN flake, which are designed during device fabrication. We outline the optimal conditions for device operation and colour centre stability, focusing on minimising background emission and ensuring prolonged operation. Our findings follow up on the existing literature and mark a step forward towards the integration of hBN based colour centres into quantum photonic technologies.

**Introduction**

Solid-state crystals are an attractive platform for advancements in quantum technologies[1,2], as they host crystal lattice defects that function as quantum emitters[3,4]. Defects formed within wide band gap materials may introduce mid-gap energy levels that are isolated from the electronic bands. These levels consist of excited and ground states that can be optically addressed to generate narrowband single photons, even at room temperature[5]. Wide band gap materials also enable site-specific creation of colour centres with reproducible zero-phonon line (ZPL) transitions[6–8], corresponding to electron decay from the excited state to ground state of the isolated two-level system. These properties, combined with various optical protocols, make such systems suitable for applications in quantum communication[4,9,10], electric and magnetic field sensing[11–13], and qubit generation[14].

For integration into on-chip quantum photonic circuits, electrical excitation of colour centres is desired[15,16]. This is typically realised in materials with well-established doping techniques for the creation of p-n junctions[17–19]. hBN, as a van der Waals (vdW) material, is particularly suitable for integration into on-chip quantum photonic circuits, enabling the realization of compact photonic circuits[20]. However, the large bandgap (~6 eV) of hBN causes challenges for p- or n- type doping. Instead, hBN offers an alternative pathway to colour centre EL, leveraging the layered nature of vdWs materials and exploiting quantum mechanisms for charge injection. Furthermore, hBN has been a well established platform host of bright, high-purity colour centres spanning the visible to near-infrared wavelength range[21–23]. Yet, their electrical excitation is still in its infancy and has only seen progress in recent years due to the complexity of charge injection and stable colour centre excitation. Few studies applied different approaches of charge injection in the hBN layer for electrical excitation of its colour centres[24–26]. Thus, it is critical to further investigate the properties of electrically generated hBN luminescent defects. Importantly, precise control of generation of electrically pumped colour centres using an electric field could enable the spatially deterministic creation of defects within a few-nanometer scale through less destructive methods through more advanced electrode engineering in future studies.

In this work, we fabricate quantum tunnelling vdW devices to enable charge injection into hBN layers. We perform a series of experiments to identify the electric field strengths required for the generation of colour centres and conduct electrical characterisation of the devices to quantify changes in the electrical properties of hBN devices. Time resolved analysis highlights spontaneous appearance of luminescent colour centres as well as charge trap generation at high applied electric fields. Generated emitters displayed very narrowband ZPL transitions at cryogenic temperatures with their wavelength mapped as a function of applied electric field. Additionally, we conduct a detailed saturation analysis to highlight the importance of optimising the device operation to minimise background emission and ensuring prolonged stability of both colour centres and devices.

**Results**

hBN EL devices were fabricated using a dry transfer technique, as detailed in the methodology section. The devices consisted of an emissive layer hBN with a thickness ranging from 20 to 40 nm, sandwiched between monolayer graphene sheets forming vertical devices. An electric field was applied across graphene layers to enable charge carrier injection, electrical generation and excitation of colour centres within the hBN layer. Figure 1a illustrates the structure of the hBN EL device, highlighting the applied electric field that facilitates charge tunneling and the excitation of luminescent colour centres present in the hBN. As discussed in the previous works[25,27–29], injection of both charges in hBN devices with low impurity density mainly follows the Fowler-Nordheim tunnelling mechanism[30] as indicated in equation (1) at cryogenic temperatures.

$$I = \frac{A_{eff} q^3 m V^2}{8\pi h \Phi_B m^* d^2} exp[-\frac{8\pi\sqrt{2m^*}\Phi_B^{\frac{3}{2}} d}{3hqV}] \quad (1)$$

Where $I$ is tunnelling current, $V$ is bias voltage, $A_{eff}$ is the effective tunneling area, $q$ is elementary charge, $m$ is free electron mass, $\Phi_B$ is tunnelling barrier height for charge carriers at the electrode and the insulator interface, $h$ is the Planck's constant, and $m*$ is the effective mass of charge carriers in the insulator. A linear relationship between $ln(I/V^2)$ and $1/V$ indicates the Fowler-Nordheim tunnelling mechanism when the equation is further formulated. Further details on this analysis are provided in the supplementary section SI1. We note that after operating the devices under the high-field regime above the stable region for an extended period and generating numerous defects, the I-V characteristics undergo significant changes and follow a defect-related tunneling mechanism rather than the pristine Fowler-Nordheim tunnelling behaviour[31–34].

This study focuses on two distinct operating regimes of the hBN EL device: low and high electric fields. In the high-field regime (≳0.75 V.nm$^{-1}$), electrical charge injection generates hBN colour centres and electrically excites them for a short period of time (within 1 hour). The low-field regime allows for prolonged excitation (several hours) of the generated luminescent defects. Generation of colour centres may arise from either the formation of new defects or changes in charged states of other pre-existing defects via electrical doping within the hBN emissive layer. This behavior is schematically depicted in Figure 1b, where the low-field regime is represented by the excitation of only few pre-existing colour centres, while the high-field regime shows the appearance of EL of additional colour centres alongside continued excitation of the original centres. Notably, non-luminescent defects are also formed during this process.

The current-voltage (I-V) characteristics of the device were measured up to approximately ~µA and are summarised in Figure 1c. Green curve represents the pristine hBN:C I-V behaviour during the initial device operation, prior to the formation of observable defects. Purple curve represents I-V characteristics after colour centres generation at 10K. Upon observing colour centres generated by electric fields, turn on voltage — bias at which current begins to increase drastically via field emission in the low field regime — produces an observable shift. This shift can be attributed to charge trap generation or the filling of additional defect states[35,36], suggesting that the charge-related properties of the device have been altered as a result of the electrical interactions between defects within the device. However, the device retains the I-V characteristics to the pristine hBN:C I-V, meaning that it continues to operate similarly despite the formation of low density defects.

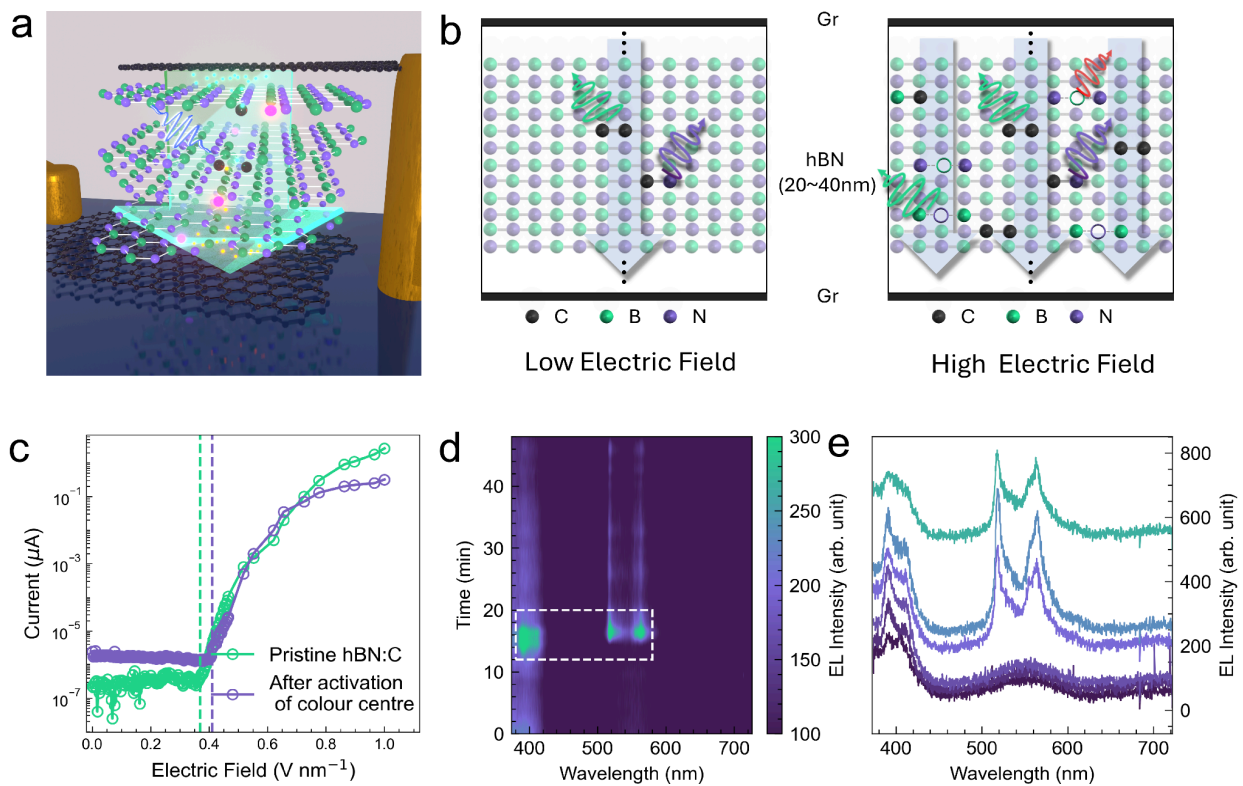

*Figure 1. Van der Waals device illustration, electrical characterization and room temperature colour centre EL spectra.* a) Three dimensional illustration of hBN colour centre EL device. b) Schematic of operation of hBN EL devices. Low applied electric field excites pre-existing colour centres within the hBN layer (left), while high electric field generates new colour centres and charge traps as well as producing EL from same and newly generated colour centres. c) Typical I-V

*characterization behaviour of a device prior and post colour centre generation. d) Time evolution of a device operated in a high field regime (0.83 V.nm$^{-1}$) at room temperature. The highlighted white dashed line shows the initial appearance of a high energy emission peak that within a few minutes switches to a stable lower energy emission. Spectra of the highlighted transition section is shown in (e) and offset vertically for clarity.*

Next, by using a scanning confocal microscope we are able to map EL photon counts intensity when the device is operated in both low and high electric field regimes. Supplementary section SI2 contains optical microscope image as well as confocal microscope EL intensity map of one of the devices. From Figure SI2b, low photon counts on the detector originate from well defined tunnelling regions that are primarily at graphene edges overlaps and the surrounding area where carriers can diffuse due to the current spreading. The bright luminescence region confined to a single confocal spot corresponds to a locally positioned colour centre that is electrically excited. Within this single confocal spot, if the device is operated at high electric field, over time newly generated luminescent defects spontaneously appear and slowly degrade in emission intensity. We studied the time-resolved EL appearance of hBN colour centres. For example, Supplementary Information Figure SI3 shows that after some time has elapsed since the device was turned on, a colour centre peak emerged at 490 nm. Over time, peak intensity decreased, because the device was still operated in a high-field regime. Further in this manuscript we show that it is possible to extend electrical excitation of generated colour centres up to 36 hours if the electric field strength is lowered post generation. Interestingly, these luminescent defects can also exhibit dynamic behavior, emitting higher-energy photons initially before undergoing a permanent rapid shift to a lower-energy transition. This phenomenon is illustrated in Figure 1d,e. Under high electric field operation, the hBN EL device initially exhibits negligible emission. After approximately 15 minutes, a distinct peak emerges at 390 nm, corresponding to high-energy emission. Within four minutes, the emission transitions to a lower-energy ZPL centred at 517 nm. Figure 1d presents spectra acquisitions showing this progression during high-field operation, while Figure 1e, is the highlighted section of individual spectra.

As discussed prior, application of a high electric field results in generation of colour centres that are also electrically excited. Emission wavelength distribution of these colour centres spans the visible to near-infrared range. Scatter plot representing ZPL transition wavelengths as a function of applied electric field at which the colour centres first appeared is shown in Figure 2a. Majority of electrically generated colour centres have ZPL transition wavelengths falling within 500 nm to 624 nm range. Interestingly, the electrical field generation method showed a high chance of generating colour centres with emission wavelengths in the range of 499 - 503 nm, which can occur anywhere within the range of 0.75 V.nm$^{-1}$ to just above 1 V.nm$^{-1}$. We also note that generation of colour centres is more frequent at a threshold electric field of 0.88 V.nm$^{-1}$. Tunnelling current values were not included in this analysis because tunnelling current conditions can vary drastically and are rather determined by the state of the hBN layer. Since high-field regime also forms charge traps, tunnelling current value is reduced upon generation of defects within hBN[37]. However, the emissive layer hBN is 20-40 nm thick and consists of many layers allowing for generation of multiple new different colour centres over time within the volume of a confocal focus. Throughout this period, other defects in hBN are constantly generated, continuously altering electrical properties of hBN. As a result, depending on the condition of the emissive hBN layer, tunnelling current value at which a colour centre is created varies more significantly than the electric field.

The generated colour centres commonly exhibit very narrowband emission transitions, that are often spectrometer-limited at 300 l/mm grating, as demonstrated in the spectra in Figure 2b. From the previous work, it was shown that emissions exhibit double Lorentzian profiles that consist of a ZPL and a phonon sideband[25], and can be expressed of the form:

$$L(x) = a_1 \cdot \frac{w_1^2}{(x-c_1)^2 + w_1^2} + a_2 \cdot \frac{w_2^2}{(x-c_2)^2 + w_2^2} \qquad (2)$$

Here, $a_i$ are amplitudes, $w_i$ are width parameters, and $c_i$ are central frequencies of first and second Lorentzian components, respectively. Such a profile is commonly maintained for most colour centres throughout the visible to near-infrared spectrum, regardless of the ZPL energy and is also demonstrated in this work with the obtained full-width at half-max (FWHM) values in Figure 2c,d falling below 0.21 nm. The plotted spectra in Figure 2c and Figure 2d were acquired in low and high applied electric fields, respectively, highlighting that such narrowband transitions occur in both regimes.

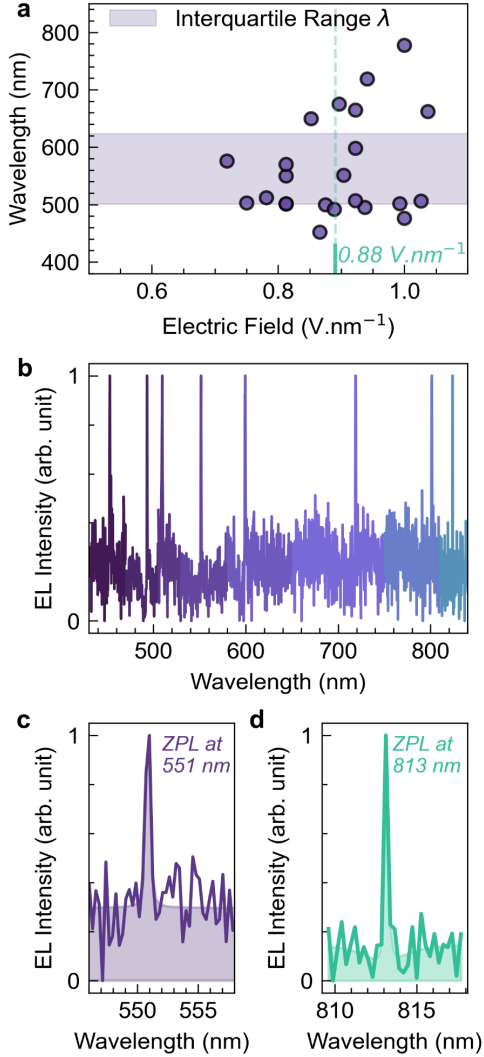

*Figure 2. Electric field parameters for generating colour centres and representative EL spectra. a) Scatter plot representing wavelength distribution of colour centres generated at specified electric field. Majority of colour centres have transitions within the 500 - 624 nm range, as shown by the interquartile range shade. Mean electric field for generating colour centres in hBN is 0.88 V.nm$^{-1}$. b) Selected representative spectra of generated colour centres spanning the whole visible to near-infrared range. c) and d) are high resolution spectra of those in (b), acquired in low and high field regimes, respectively. Double Lorentzian fits were used for fitting, yielding FWHM values of 0.17 and 0.20 nm for colour centres at 551 nm and 813 nm respectively. Measurements conducted at 10 K.*

When the devices are operated in a high-field regime, in addition to generating and electrically pumping hBN colour centres, a second type of emission that spans broadly across the visible spectral range is also produced. Figure 3a shows three representative spectra obtained from an EL device at three distinct electric field regimes, highlighting two emission components that are particularly obvious during high-field operation. We report observation of a broad emission spectrum that begins to dominate above 0.8 V.nm$^{-1}$ and likely arises from thermal emission due to current flow in tunnelling regions. In contrast, the narrowband emission is characteristic of hBN colour centres, originating from localised defect states that emit photons through electronic transitions. As the electric field increases, the narrowband emission remains spectrally distinct, while the thermal emission becomes increasingly dominant in intensity. These observations emphasize the presence of two entirely different mechanisms contributing to the emission spectra within the tunnelling region of a device. It is important to consider the broad emission as it contributes significant number of counts to the spectrum, particularly at high electric fields. This introduces challenges when performing measurements on single colour centres, such that an appropriate electric field operating regime must be carefully determined to minimise contribution of the background emission while ensuring efficient excitation of colour centres, as depicted in Figure 3.

We further investigate the two emission mechanisms by analysing their dependencies on electric field and current, represented in Figure 3b–f. The broad thermal emission (green) exhibits a strong nonlinear increase in emission intensity as a function of applied electric field, while having a more linear slope with applied current. In contrast, the narrowband colour centre emission (blue) shows a distinct saturation behaviour as a function of current that is characteristic of luminescent defects and is governed by the relationship $C = \frac{C_{\infty} I}{I + I_{SAT}}$ that yields a finite saturation luminescence counts value $C_{\infty}$, where $I$ and $I_{SAT}$ represent tunnelling current prior and at saturation, respectively. This limit is imposed by the rate of change of population of a ground energy state and the spontaneous decay of electrons from the excited state. The rate of spontaneous decay is dependent on the properties of the system that govern the lifetime of an electron in an excited state until it relaxes to a ground state via photon emission. The studied colour centre saturates at approximately 9 counts with 4.42 µA of tunnelling current flowing through the emissive hBN layer.

The semi-logarithmic plot in Figure 3f summarises the interplay between these two mechanisms. As the current increases exponentially with the electric field, thermal emission is absent until ~0.75 V.nm$^{-1}$, after which it begins to rapidly increase in counts and dominate emission spectrum contribution at fields above 0.8 V.nm$^{-1}$. Contrary, the colour centre has a lower emission turn on electric field, ~0.68 V.nm$^{-1}$ and follows a less aggressive slope. These observations further confirm that the two emission components arise from fundamentally different physical processes, with the broad-band photons likely being due to thermal emission, while the colour centre luminescence tied to quantum defect luminescence.

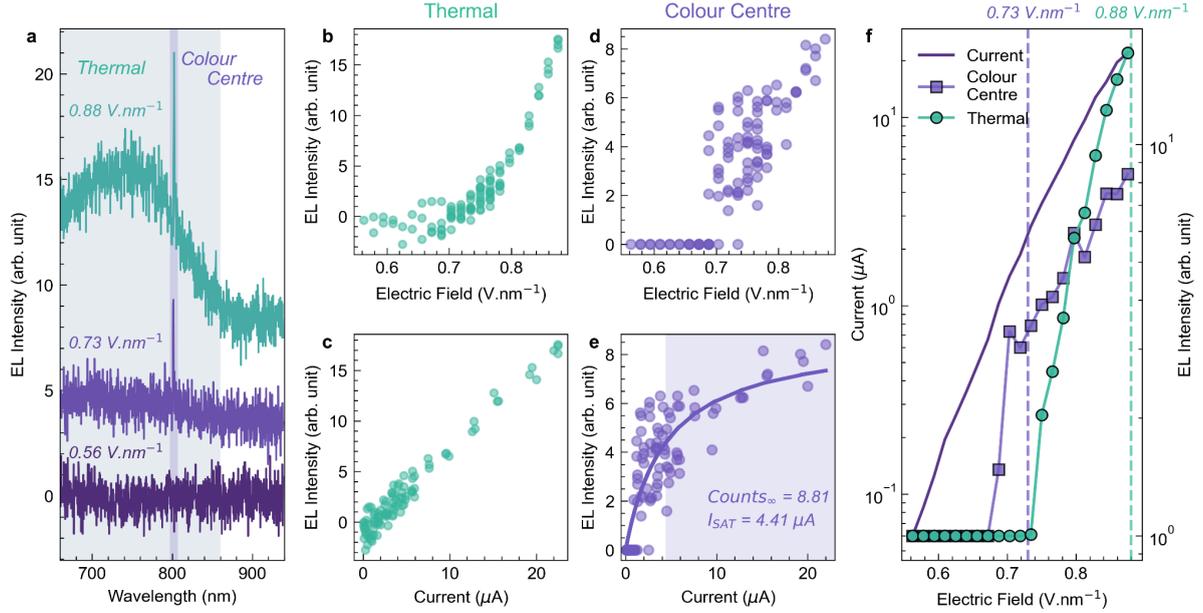

*Figure 3. Thermal and electrically driven colour centre emissions.* a) Three representative spectra taken at different electric field regimes; minimal field showing no emission, low field showing only colour centre EL and high field showing both thermal and colour centre emissions. Thermal EL counts obtained from within the blue shaded region are plotted in (b) and (c) against electric field and current, respectively. Colour centre EL counts obtained from double Lorentzian fits of the narrowband peak plotted in (d) and (e), respectively. Blue shaded region in (e) highlights data points that are past the saturation current value. f) Semi-logarithmic plot of current and EL counts of both colour centre and thermal emissions against applied electric field. Vertical dashed lines indicate the same electric fields at which representative spectra in (a) were acquired. Measurements conducted at 10 K.

We also investigated the dynamic response of colour centres and devices to applied electric fields. In these measurements, the electric field was varied over time while simultaneously recording the colour centre emission and tunnelling currents, as presented in Figure 4.

In Figure 4a, we first examine the behaviour of a colour centre over time as the electric field changes. A colour centre was generated within the emissive hBN layer prior to the measurement by operating the device at 0.75 V.nm$^{-1}$ for 10 minutes. The top panel shows the electric field variation over 14 minutes: starting at 0.69 V.nm$^{-1}$, increasing to 0.78 V.nm$^{-1}$, decreasing to 0.58 V.nm$^{-1}$, and increasing again to 0.88 V.nm$^{-1}$. The corresponding emission counts from the colour centre (bottom panel) exhibited a dynamic response to applied electric field, revealing that no luminescence was observed below a threshold field value of 0.68 V.nm$^{-1}$. We also demonstrate that the same colour centre was switched back into luminescent state if the applied electric field was above the threshold value. However, at extreme electric fields, nearing 1 V.nm$^{-1}$, the colour centre would permanently switch off its luminescence. This behaviour is highlighted past the 13.5 minute mark, when EL from this colour centre has disappeared.

Continuous increase of applied bias in the high-field regime typically resulted in permanent switching off of colour centres. Furthermore, extended operation of devices in a high-field regime also caused a similar effect, although it was often delayed. Thus, in addition to optimising operating conditions to minimise background contributions, it is also essential to map out safe electric field ranges to preserve colour centre emission. In Figure 4b, we quantify the time for which individual colour centres remained active at specific electric fields. Measurements repeated across multiple colour centres revealed that fields around 0.78 V.nm$^{-1}$ rapidly quenched their emission. In contrast, reducing the electric field below 0.72 V.nm$^{-1}$ allowed for significantly prolonged operation. Under these optimised conditions, we achieved continuous electrical excitation of a single colour centre for up to 36 hours.

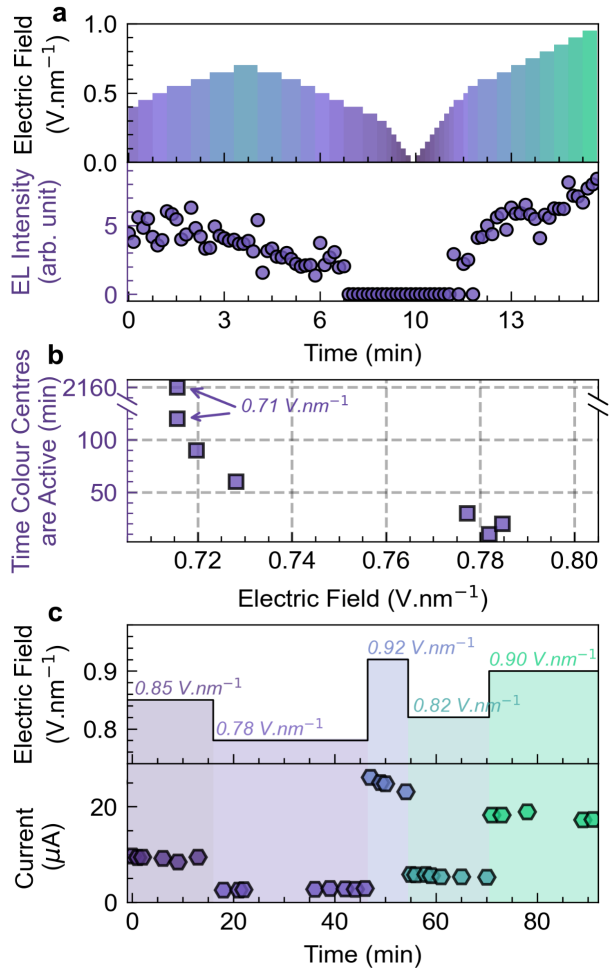

*Figure 4. Device and colour centre time dynamics over varying electric fields. a) Time dynamics of EL from a colour centre at varying electric fields over time. Colour in the top panel corresponds to electric field strength. b) Time dependence of colour centres continuously measured at specified electric fields. c) Plots of electric fields and measured currents over time during measurement of a device. Measurements conducted at 10 K.*

To further investigate the dynamic response of devices to applied electric fields, tunnelling current was monitored over time as the electric field was varied. These measurements are depicted in Figure 4c. The tunnelling current exhibited a rapid response to changes in the applied electric field, adjusting instantaneously to a new set electric field. Application of high electric field on hBN EL devices results in multitude of physical changes of the emissive layer, which could be observed via tunnelling current behaviour. At high electric fields, tunnelling current would begin to drop, likely due to generation of charge traps. This is particularly noticeable at very high electric fields, above 0.9 V.nm$^{-1}$, where severe current drop is observed within a time span of less than 10 minutes. The effect also persists at slightly lower electric fields (>0.8 V.nm$^{-1}$), but with more gradual reduction in current. In contrast, electric fields below 0.8 V.nm$^{-1}$ tend to maintain stable current values over time, in fact at 0.78 V.nm$^{-1}$, the current fluctuated around 2.7 µA for 28 minutes, demonstrating stable device operation under these conditions.

Lastly, we report the observation of room temperature photoluminescence (PL) charge state control of a colour centre under applied electric fields. While similar behavior has been studied in other solid-state colour centres[35,38], this work demonstrates a clear correlation at room temperature with a unique observation. Specifically, a device hosting a colour centre with an emission energy of 432 nm demonstrated that, in PL measurements, applied electric fields induced the appearance of an additional lower-energy peak within the 580–655 nm wavelength range. The intensities of both peaks were found to be dependent on the bias polarity, with higher counts observed under positive electric fields. Notably, the lower-energy peak further split into a doublet within the ~600 nm range when the applied positive electric field was sufficiently large. This analysis is summarised in Figure SI4.

**Conclusion**

To summarise, engineered hBN EL devices were investigated for their ability to generate colour centres under applied electric fields. This was evident through I-V, and time-resolved spectral and current analyses that showed formations of other non-luminescencent defects, such as charge traps. We highlighted the electric fields necessary for generation of visible to near-infrared wavelength range narrowband colour centres. Most importantly, we determine the best conditions of operating tunnelling devices for minimised background and prolonged operation of both colour centres and devices. Our findings conclude that generated colour centres are most optimally measured at electric field strengths around 0.7 V.nm$^{-1}$. Our work advances the ongoing progress towards on-chip integration of quantum optoelectronic van der Waals systems based on hBN.

**Methods**

Device Fabrication

The hBN, hBN:C and graphene crystals were mechanically exfoliated onto a 90nm oxide layer and thickness characterized by atomic force microscopy (AFM). As an emissive layer hBN, we used hBN:C within the 20 to 40 nm range. Carbon related defects were introduced into high purity hBN crystals by annealing in a graphite furnace at 2000 °C. Exfoliated hBN:C crystals were additionally annealed under Ar conditions (950 °C, 50 sccm, 1.3 mbar). The vdW heterostructures were fabricated using dry transfer techniques. Polyethylene terephthalate (PET) is utilized as a stamp, stacking hBN/Gr/hBN/Gr/hBN heterostructures. Each layer is precisely aligned by a home-built transfer stage based on an optical microscope. Picking up each layer was conducted at 70 °C step by step, and released the whole structure onto a Ti/Au pre-patterned 300 nm thick SiO2/Si substrate and heated up to 130 °C to release. The PET stamp was removed in dichloromethane for 12 h at room temperature and 1 h post heat treatment at 60 °C. Electrode patterns to connect graphene to pre-patterned electrodes were used and Ti (10 nm) and Au (40 nm) were evaporated to generate an elaborate pattern.

Optical Measurements

For room temperature measurements, devices were kept under high vacuum conditions, approximately 10$^{-5}$ mbar. Cryogenic temperature measurements were conducted using a closed-loop helium cryostat. The devices were connected to a sourcemeter (Keithley 2400 series) for application of electric field and the current value was monitored in real time with GPIB. High NA objectives (>0.8 NA) were used for collecting light from the samples. Home-built confocal microscope systems coupled the collected light into multimode fibers for either confocal raster scan with avalanche photodiodes (Excelitas APDs) or imaged on spectrometer (Andor SR303I-Newton CCD) using 300 l/mm grating.

**Acknowledgements**

We acknowledge financial support from the Australian Research Council (CE200100010, FT220100053, DP240103127), and the UTS node of the ANFF for access to nanofabrication facilities. K.W. and T.T. acknowledge support from the JSPS KAKENHI (Grant Numbers 21H05233 and 23H02052) and World Premier International Research Center Initiative (WPI), MEXT, Japan. G.P., and J.K. acknowledge the support from Institute of Basic Science (IBS-R034-D1). J.K. acknowledge the support from the National Research Foundation of Korea grants (NRF-2023R1A2C2007998). This study is supported by Brain Korea 21 FOUR project for Education and research centre for future materials (F21YY7105002). This study was also supported by the MSIT (Ministry of Science and ICT), Korea, under the ITRC (Information Technology Research Center) support program (IITP-2023-RS-2022-00164799) supervised by the IITP (Institute for Information & Communications Technology Planning & Evaluation).